\documentclass[sn-mathphys]{sn-jnl}

\jyear{2021}%

\theoremstyle{thmstyleone}%
\newtheorem{theorem}{Theorem}
\theoremstyle{thmstyletwo}%

\theoremstyle{thmstylethree}%

\def\be{\begin{equation}}
\def\ee{\end{equation}}
\def\bea{\begin{eqnarray}}
\def\eea{\end{eqnarray}}
\def\bean{\begin{eqnarray*}}
\def\eean{\end{eqnarray*}}

\usepackage{bm}

\begin{document}

\title[Penrose's influence in General Relativity]{The influence of Penrose's singularity theorem in General Relativity}


\author*[1,2]{\fnm{Jos\'e M. M.} \sur{Senovilla}}\email{josemm.senovilla@ehu.eus}

%

\affil[1]{\orgdiv{Departamento de F\'{\i}sica}, \orgname{University of the Basque Country UPV/EHU}, \orgaddress{\street{Apartado 644}, \city{Bilbao}, \postcode{48080}, 
\country{Spain}}}

\affil[2]{\orgdiv{EHU Quantum Center}, \orgname{University of the Basque Country UPV/EHU}
}



\abstract{Penrose's crucial contributions to General Relativity, symbolized by his 1965 singularity theorem, received (half of) the 2020 Nobel prize in Physics. A renewed interest in the ideas and implications behind that theorem, its later developments, and other Penrose's ideas improving our understanding of the gravitational field  thereby emerged. In this paper I highlight some of the advancements motivated by the theorem that were developed over the years. I also identify some common misconceptions about the theorem's implications. A modern perspective on the concept of closed trapped submanifolds, based on the mean curvature vector, is advocated. 
}




\maketitle

\section{Introduction}\label{sec1}
The importance of Penrose's contributions \cite{Pen,P0,P00,P,P2,P23,P3,P4,P5,PI,P6,Pextra,PR1,PR2} to the understanding of the gravitational field and the global structure of spacetime is enormous. In particular, his 1965 singularity theorem \cite{P,SMilestone} was a gigantic milestone in the history of General Relativity (GR) that changed the field for ever more. As argued elsewhere \cite{S5,SMilestone}, this result genuinely started the post-Einsteinian era of GR.

The beautiful and fundamental concept of {\em closed trapped surface} was introduced as the key assumption to prove the theorem in \cite{P} and most of the many singularity theorems that came later \cite{HE,HP,S1}. I will devote a substantial part of this contribution to this idea from a modern perspective, and to highlight some of its less-known properties as well as to the improvements that are currently fashionable. I will also discuss some common misconceptions about the implications of the singularity theorem, and discuss its shortcomings and strengths.

However, Penrose's impact onto GR is much deeper and wider than the singularity theorem and includes many other wonderful and fruitful ideas, such as using the spinorial approach \cite{Pen,NP}, conformal treatment of spacetime \cite{P0,P00,P23,P3}, causal properties of spacetime and the causal boundary \cite{P5,KrP,GKP}, asymptotic conserved quantities \cite{NP1,NP2,ENP}, gravitational waves and their collision \cite{P2,KhP}, black holes \cite{P4,PI,PF} and, of course, the question of --weak and strong-- cosmic censorship \cite{PI,P6,P7} (the list is far from complete!).

I will also dwell into these subjects, especially for those advencements arising from the 1965 theorem, presenting some (hopefully) interesting remarks that are not usually considered in the standard literature. A notable exception will be cosmic censorship that I will just discuss tangentially herein and for which I refer to \cite{Wald1} and to Landsman's contribution to this Special Issue \cite{landsman} where the subject is masterly discussed. 

\section{The 1965 theorem, and a 2022 version}\label{sec:Th}
The original Penrose ``singularity'' theorem is actually an incompleteness theorem and contained the following assumptions
\begin{enumerate}
\item A 4-dimensional Lorentzian manifold with a metric $g$ of class $C^2$ and causally oriented. 
\item a non-compact $C^3$ Cauchy hypersurface $\Sigma$
\item a closed future-trapped surface 
\item $R_{\mu\nu} k^\mu k^\nu \geq 0$ for all null $k^\mu$ (null convergence condition, $R_{\mu\nu}$ is the Ricci tensor)
\item all null geodesics are complete towards the future of $\Sigma$
\end{enumerate}
The proof of the theorem showed that these 5 hypotheses are inconsistent. Hence, by following the traditional reformulation of the theorem \cite{HP,HE}, one usually states it by saying that, if 1--4 hold, then spacetime must have future-incomplete null geodesics. Observe that the theorem is independent of the Einstein field equations
\be\label{eq:EFE}
R_{\mu\nu}-\frac{1}{2}R\,  g_{\mu\nu}+\Lambda g_{\mu\nu}=\frac{8\pi 
G}{c^4}T_{\mu\nu}
\ee
where $R=R^\mu {}_\mu$ is the scalar curvature, $G$ is Newton's gravitational 
constant, $c$ is the speed of light in vacuum and $\Lambda$ the 
cosmological constant. 

Let me scrutinize these 5 assumptions and their meaning.

\subsection{Incompleteness. Causal boundary. Isocausality}
Consider first hypothesis 5 whose negation, as just mentioned, became the standard {\em conclusion} of the general singularity theorems \cite{HE,S1,SMilestone}. The main problem with what relativists want to consider {\em singularities} is that spacetime breaks down at them, so that singularities are not part of the spacetime \cite{Ge2}. Still, physical curves are good pointers and they do belong to the spacetime. By using curves that are maximally extended to the future, say, (that is, no further continuation of the curve is feasible to the future) such curves may be complete or not. In the former case they go up to infinity in the future, in the latter case the canonical parameter (affine parameter for geodesics) only achieves a finite value and they are pointing towards a problem within reach: the singularity. 
Under the influence of Hawking \cite{H0,H00,H1,H2,H3,HE} and Geroch \cite{Ge-,Ge2} (geodesic) incompleteness eventually became the standard characterization of singularities proven by the theorems.

Actually, the ideas contained in the previous paragraph are the fundaments for the construction of the {\em causal boundary} \cite{GKP} assuming some basic causal properties, such as a distinguishing spacetime (for the causal ladder and more on causality see \cite{MSa,GS,Mi1}). Basically, instead of working with points in spacetime one deals with causal curves, and their pasts and futures. If a curve terminates (to the future) at a given point, then the past of the causal curve and that of its endpoint are the same. Similarly, {\em mutatis mutandis}, for curves that start from a point in the past. Thus we can replace points by pasts and futures of curves. Obviously, some identifications are required, but these technicalities can usually be resolved satisfactorily. Now, to this set of pasts and futures we can add the future and past sets that are {\em not} the future or past of any point. They correspond to futures and pasts of {\em inextendible} causal curves, and are then considered to be the elements of a new, larger, set which contains all the points of the spacetime plus ``points on the boundary''. Basically, this defines the causal boundary as those points on the larger set that are not points of the original manifold. The causal boundary contains both points at infinity and {\em singularities} --as also does the conformal boundary \cite{Fra,P00,P3,HE}. For the history and development of these important concepts I refer to \cite{GS,Flo}. For the most advanced and up-to-date constructions, see \cite{Flo,FH,FHS0,FHS,FHS1}.

The ideas of the causal boundary are based on the more fundamental {\em causality theory}. Classical references are \cite{HE,P5,S1} but more recent, more complete accounts are given in \cite{GS,MSa,Mi1} and some references therein. In particular, one can consider abstract definitions of causal spaces \cite{KrP,Ca} which go beyond the scope of GR. For the case of a Lorentzian manifold, the whole causality theory is based on the fundamental result that states that in any normal neighbourhood of any point $p$ the exponential map defines a light-cone on the manifold (as the image of the null cone in $T_p M$) such that null geodesics are precisely on that light cone and any other causal curve lies in the interior of the cone from the point that it fails to be a null geodesic on \cite{BEE,Ca,HE,Kri,S1,GS}. This basic result can actually be improved in many ways by lowering the differentiability of the metric and other requirements \cite{Mi1}. Sometimes, this result is interpreted as the local equivalence of causality with that of flat spacetime. Unfortunately, this is an erroneous belief {\em if causal equivalence is assumed to mean conformal equivalence}, as it is customary. This question was addressed in \cite{KrP} where the authors proved, by providing a counterexample, that the local equivalence of causal (conformal) structure between different Lorentzian manifolds is simply a chimera: it does not hold. 

The definition of local causal equivalence that they used was as follows (see also the discussion in \cite{GS}):

{\it For each point $p$ of the Lorentzian manifold $(M,g)$ there exists a small open neighbourhood $U_p$ of $p$ and a homeomorphism $h_p$ of $U_p$ onto an open subset of Minkowski spacetime such that $h_p$ is a causal isomorphism.}

Here, assuming that $(M,g)$ is distinguishing \cite{MSa,S1,Mi1}, a causal isomorphism is actually a smooth conformal isometry \cite{GS}. The failure of this ``natural'' definition is an unsatisfactory situation that can be remedied by adopting a better definition of causal equivalence introduced in \cite{GS0}; {\em isocausality}. Two spacetimes $(M,g_1)$ and $(M,g_2)$ are said to be {\em isocausal} if the there exist (i) a diffeomorphism $\Phi$ that sends every future-pointing vector in $(M,g_1)$ to a future-pointing vector in $(M,g_2)$ and (ii) another diffeomorphism $\Psi$ that sends every future-pointing vector in $(M,g_2)$ to a future-pointing vector in $(M,g_1)$. The important point here is to realize that $\Psi$ does not necessarily have to be $\Phi^{-1}$. Actually, $\Psi=\Phi^{-1}$ only when $g_1$ and $g_2$ are conformally related. Isocausality preserves most of the causal properties of Lorentzian manifolds \cite{GS0,GSa} though may not preserve the causal boundary \cite{FHS0}.

With the concept of isocausality, the above characterization of local causal equivalence can be modified as \cite{GS}:

{\it For each point $p$ of the Lorentzian manifold $(M,g)$ there exists a small open neighbourhood $U_p$ of $p$ and an open subset ${\cal U}$ of Minkowski spacetime such that $U_p$ and ${\cal U}$ are isocausal.}

Using this new definition, any two Lorentzian manifolds (of the same dimension) are locally isocausal \cite{GS}.

\subsection{Cauchy hypersurfaces. Global hyperbolicity}\label{subsec:Cauchy}
Consider now hypothesis 2, existence of a non-compact Cauchy hypersurface. A {\em partial Cauchy hypersurface} $\Sigma$ is any edgeless closed acausal set. Acausality means that no pair of points in the set are causally related. For the definition of edge see \cite{HE,S1}, it basically states that it has no end. The Cauchy development or {\em domain of dependence} \cite{Ge3} of $\Sigma$, denoted by $D(\Sigma)$, is given by all points $x$ such that either every future-directed, or every past-directed, inextendible causal curve from $x$ meet $\Sigma$. $\Sigma$ is called a global Cauchy hypersurface, or simply a Cauchy hypersurface, if its domain of dependence is the entire manifold, that is, $D(\Sigma)=M$.

One usually considers {\em spacelike} Cauchy hypersurfaces, though there can be other ones. Therefore, in simple words, a Cauchy hypersurface is a spacelike slice amenable to sustain initial data that can determine the full spacetime completely. This is of course the basis for the initial value formulation of the GR field equations \cite{Choquet,ChGe} , see \cite{Ringstrom} for details. One can prove \cite{Ge3} that $\Sigma$ is a Cauchy hypersurface if and only if every inextendible null curve crosses it exactly once (and then, every timelike inextendible curve also intersects $\Sigma$ once). Spacetimes with Cauchy hypersurfaces are called {\em globally hyperbolic} \cite{HE,P5,Wald}, a concept that comes from previous works by Leray \cite{Le}, because the sets containing all causal curves between any two points are compact --and therefore, there will always be a maximizing geodesic between any two causally related points. 

Globally hyperbolic spacetimes possess the best causal behaviour \cite{GS,MSa}, and therefore they are at the top of the causal ladder. From a physical point of view, it seems reasonable to believe that realistic spacetimes will be globally hyperbolic. Actually, the strongest versions of cosmic censorship aspire to prove this under very general circumstances \cite{landsman}. As proven originally by Geroch, globally hyperbolic spacetimes have a fixed spatial topology (given by that of any Cauchy hypersurface $\Sigma$) as they are topologically $M=\mathbb{R}\times \Sigma$ and actually all spacelike Cauchy hypersurfaces are diffeomorphic, see \cite{BeSa,BeSa1,BeSa2,Mi1} for the latest on this subject.

The concept of domain of dependence and of global hyperbolicity is very powerful. As an interesting example, consider the exterior region of the vacuum Schwarzschild solution, given in standard coordinates by
\be\label{Schw}
ds^2 = -\left(1-\frac{r_g}{r}\right) c^2 dt^2 + \frac{dr^2}{1-\frac{r_g}{r}} +r^2 \left( d\theta^2 +\sin^2\theta d\varphi^2\right)
\ee
with $t\in (-\infty ,\infty)$, $\theta\in [0,\pi)$, $\varphi \in [0,2\pi)$ and $r\in (r_g,\infty)$, where the {\em gravitational} or {\em Schwarzschild} radius $r_g$ is a positive constant. This metric is globally hyperbolic, any $t=$ const.\ slice being a Cauchy hypersurface. This implies, {\em without performing any extension nor any further calculations} that $r\rightarrow r_g$ must be a null boundary, and therefore this metric has not the structure of Minkowski minus a cylinder of radius $r_g$. This is a very simple way to learn that $r=r_g$, not being a curvature singularity, is actually a horizon without having to perform any extension of the metric or other calculations, see \cite{GS0} for further comments. 

Coming back to the theorem's hypothesis 2, assuming that $\Sigma$ is non compact amounts to saying that the spacetime is open, that is to say, space is not finite. Actually, according to Minguzzi \cite{Mi} the assumption of the existence of $\Sigma$ can be considerably relaxed, so that the main result is kept even in absence of total predictability. Still, the assumption of spacetime being ``spatially open'' was needed in \cite{Mi}, so that the idea that space is infinite survives. Whether or not this is reasonable is dabatable, but in any case it looked like a reasonable assumption for the analysis of spacetimes with collapsing matter. 

There are well-known examples of geodesically complete spacetimes without Cauchy hypersurfaces. A dramatic example is also due to Penrose in 1965, {\em plane waves} \cite{P2}. Plane waves are solutions of the Einstein-Maxwell equations (including the vacuum case) without cosmological constant\footnote{Vacuum or electromagnetic plane waves' Ricci tensor takes the form $R_{\mu\nu}= -(A(u)+C(u))  k_\mu k_\nu$ where $k^\mu$ is the parallel vector field of the spacetime} with a parallel null vector field $\vec k = \partial_v$ orthogonal to null hyperplanes $\bm{k} =-du$. This vector field defines the direction of the propagating waves. Using coordinates $x,y$ on the planes orthogonal to $\bm{k}$ the line-element can be written as
\be \label{PW}
ds^2 = -2du dv +dx^2 +dy^2 + \left(A(u) x^2 +2B(u) x y + C(u) y^2 \right) du^2
\ee
where $A,B$ and $C$ are three arbitrary functions of the retarded time $u$. The non-zero components of the Riemann tensor in the coordinate basis are
$$
R_{uxux} = - A , \hspace{1cm} R_{uxuy} =-B , \hspace{1cm} R_{uyuy} =-C
$$
so that there are no curvature singularities whatsoever as long as the three functions are regular everywhere. Under these conditions, the geodesic completeness of these spacetimes was proven in \cite{FS}. Plane waves cannot contain closed trapped surfaces \cite{MS,S2003}, and this will be briefly discussed later in subsection \ref{subsec:trapped}. 

It is quite remarkable that these spacetimes, that describe appropriately physically realistic gravitational waves far enough from the sources, are free from singularities and complete, nevertheless, failing to be globally hyperbolic, they are not {\em strongly} cosmic censored. This may seem controversial, and is basically due to the structure of its causal boundary (\cite{GS} and references therein) but that is how things are at present, see \cite{landsman}.

A question naturally arises: what about globally hyperbolic {\em and} geodesically complete spacetimes? There are several well-known examples, such as (i) the Einstein static universe metric ($\Lambda >0$) \cite{E0}
$$
ds^2= -c^2 dt^2 + \frac{1}{\Lambda} \left[d\chi^2 +\sin^2\chi \left(d\theta^2 +\sin^2\theta d\varphi^2\right)\right]
$$
that satisfies (\ref{eq:EFE}) for a dust $T_{\mu\nu} = \varrho u_\mu u_\nu$ with $\mathbf{u}=-cdt$, $8\pi G \varrho = 2c^4 \Lambda$ and $\chi\in(0,\pi)$ so that the 3-dimensional metric in parenthesis is the standard metric of a round 3-sphere. (ii)  de Sitter spacetime \cite{Exact,GP} given by
$$
ds^2= -c^2 dt^2 + \lambda^2\cosh^2(ct/\lambda) \left[d\chi^2 +\sin^2\chi \left(d\theta^2 +\sin^2\theta d\varphi^2\right)\right], \hspace{7mm} \lambda^2=3/\Lambda
$$
where, again, the $t=$const.\ 3-spheres are Cauchy hypersurfaces with the standard round metric. This is a vacuum solution of (\ref{eq:EFE}) with $\Lambda>0$: $T_{\mu\nu}=0$. (iii) any global spherically symmetric spacetime obtained by properly matching an interior solution to the Schwarzschild exterior (\ref{Schw}) across an $r=$const.\ hypersurface. (iv) A less symmetric and dynamical example is given by $(\mathbb{R}^4,g)$ in cylindrical coordinates $\{t,\rho,\varphi,z\}$ with line element \cite{S}
\begin{eqnarray}
ds^{2}&=&\cosh^4(at)\cosh^2(3a\rho)(-c^2dt^2+d\rho^2)+\cosh^{-2}(at)\cosh^{-2/3}(3a\rho)dz^2 \label{sol}\nonumber\\
&+&\frac{1}{9a^2}\cosh^4(at)\cosh^{-2/3}(3a\rho)\sinh^2(3a\rho)d\varphi^2 \nonumber
\end{eqnarray}
where $a$ is a positive constant. This metric satisfies (\ref{eq:EFE}) with $\Lambda=0$ for a perfect fluid in comoving coordinates
with energy density 
$$
\frac{8\pi G}{c^4}\varrho = 15a^2\cosh^{-4}(at)\cosh^{-4}(3a\rho)
$$
and a radiation dominated equation of state 
$$
p=\frac{1}{3}\varrho .
$$
For proofs of the geodesic completeness and the global hyperbolicity of this solution see \cite{CFS}.
A discussion about how these examples avoid the incompleteness predicted by the most powerful singularity theorems can be found in \cite{CFS,S6}.

The previous list contains just (standard or well-known) {\em particular }examples. These examples provide a lot of information about geodesically complete and globally hyperbolic spacetimes, but one wonders if general results can be found while keeping the convergence condition
\be
R_{\rho\nu}v^{\rho}v^{\nu}\geq 0 \label{ConCon}
\ee
for arbitrary causal $v^\mu$. The answer is yes. 

First, in the stationary case, if (\ref{ConCon}) holds then geodesic completeness requires \cite{GaH}
$$
\frac{R_{\mu\nu}\xi^{\mu}\xi^{\nu}}{\xi^{\mu}\xi_{\mu}}\sim k/\bar\rho^2
$$ 
for some constant $k$, where $\vec\xi$ is the timelike Killing vector field while $\bar\rho$ is an appropriate spatial distance between any two events. 

The dynamical and `closed universe' situation is not possible if the model is expanding (or contracting), as standard singularity theorems apply \cite{HE}: closed expanding non-singular models necessarily require the violation of (\ref{ConCon}), see \cite{S1}.

In the dynamical and `open universe' case, for expanding (or contracting) models, if (\ref{ConCon}) is satisfied then at least one of the following must hold to ensure geodesic completeness \cite{SRay,S3} 
\begin{itemize}
\item the cosmological constant $ \Lambda <0$ is negative
\item the \underline {averaged} energy density on a Cauchy hypersurface vanishes
\item the \underline {averaged} intrinsic scalar curvature of a Cauchy hypersurface vanishes
\end {itemize}
Here `averaged $X$' means the variable $X$ integrated over a Cauchy hypersurface $\Sigma$ divided by the volume of $\Sigma$, or the appropriate limit thereof if $\Sigma$ is non-compact \cite{SRay,S3}. The conclusion is thus similar to that of the stationary case: the energy density at any instant of time must fall off quickly enough in spatial directions.

Dynamical situations where the world is expanding somewhere but contracting somewhere else are still open for study.

\subsection{Spacetime dimension. Differentiability of the metric}\label{subsec:dim}
Let me analyze now assumption 1. It has three different parts, the causal orientation, the 4-dimensional Lorentzian manifold, and the $C^2$ differentiability of the metric. With regard to the first, this is just assuming that there is a well-defined and consistent future direction in the spacetime. Basically it requires the existence of a global smooth vector field in the manifold.

Concerning the dimensionality, actually Penrose proved the theorem just in 4 dimensions for obvious reasons, but it holds in arbitrary dimension $n$ greater than or equal to 3. Of course, this requires some minor adjustments, basically that the closed trapped surface is a compact (without boundary) spacelike submanifold of dimension $n-2$; equivalently, that its co-dimension is 2. Of course, the dimensionality of the Cauchy hypersurface is always $n-1$, a co-dimension 1 submanifold in any $n$. Yet, when analyzing the more powerful Hawking-Penrose singularity theorem \cite{HP,HE}, the question arises of why the {\em boundary assumption} in the theorem requires either a hypersurface, a trapped surface, or a point with re-converging light cones. That is, the boundary hypothesis is placed on submanifolds of co-dimension 1, 2 or $n$. Is there anything wrong if one wishes to place a boundary condition in submanifolds of co-dimension $3,\dots, n-1$? The answer is actually negative, and this will be discussed and clarified at large in subsections \ref{subsec:trapped} and \ref{subsec:EnergyCond}.

With regard to differentiability, the traditional singularity theorems required the metric to be at least of class $C^2$. This is actually needed in many intermediate steps necessary for proving the theorems. A list of these places can be found in section 6.1 of \cite{S1}. At first sight, it may seem reasonable to deal with $C^2$ metrics, however, this would leave out important metric models that are built by matching spacetimes \cite{MS0}, such as the gravitational field of a spherical star or shock (electromagnetic or gravitational) waves, among others. 

For a long time, this excessive assumption of differentiability was one of the shortcomings of singularity theorems, because one might think that actually the geodesic incompleteness would not obtain, but rather that the metric could just be of class $C^1$. This should not count as a true singularity, especially if geodesics can be uniquely continued. After such a long waiting time, things starting to change with the results in \cite{Min15,KSSV14} where local results concerning convex normal neighbourhoods and the existence, uniqueness and properties of geodesics were found for metrics of class $C^{1,1}$, that is, with first partial derivatives Lipschitz continuous. It must be observed that the Riemann tensor is not continuous in this situation, having finite jumps. Thus, it is not defined pointwise, and the assumptions and strategies in the theorems must be adapted accordingly. Despite these difficulties, soon after, the first singularity theorem to be proven in $C^{1,1}$ regularity \cite{KSSV15} was the classical one of \cite{H3}, and Penrose's theorem came immediately after \cite{KSV}. Nevertheless, the goal to prove the more demanding Hawking-Penrose theorem \cite{HP} required some extra effort and even new techniques, but after some 3 more years the goal was achieved in \cite{GGKS}.

From a physical point of view, such a result may seem satisfactory and no further relaxation of differentiability looks desirable in principle. For pure $C^1$ (or lower) metrics the curvature tensor is not well defined as a tensor field and one must resort to the use of tensor distributions. Still, there are important idealizations of physical situations with such type of distributional curvature, such as {\em impulsive gravitational waves}. Actually, the very first collision of plane gravitational waves, due to Khan and Penrose \cite{KhP}, used such impulsive waves. And by the way, the collision of plane waves is known to lead to singularities in generic situations \cite{Gri,T7,Y2}. Therefore, the quest for incompleteness theorems with even lower regularity continued, and the key results were put forward in \cite{Graf}. Extra effort, once again, has been needed to prove the Hawking-Penrose theorem for $C^1$ metrics, but the result is already claimed to be achieved recently \cite{KOSS}.

For an up-to-date summary of all these advances lowering the differentiability of the metric, detailed explanations and future perspectives, check \cite{stein}.

\subsection{Trapped submanifolds}\label{subsec:trapped}
Consider now hypothesis 3. This is the most important legacy of Penrose's singularity theorem, a brilliant and prolific idea with innumerable applications, see e.g. \cite{SMilestone} for an account. 

In GR the geometry, the curvature, of space–time is the gravitational field and therefore basic
geometrical quantities --such as lengths, areas or volumes--- will undergo a time evolution in generic dynamical scenarios. A closed (future) trapped surface appears when its area instantaneously decreases (to the future) no matter how they evolve in a causal way. This is why they are called trapped --though perhaps trapping would be better--, for everything they contain will be surrounded by surfaces of less area. There is a dual version to the past. This idea cleverly captures the concept of ‘point of no return’ in, for instance, stellar gravitational collapse.

Of course, this can be mathematically well defined for arbitrary submanifolds of any dimension $n-m$ (co-dimension $m$). From classical geometry we know that the instantaneous `evolution' of the $m$-volume of any submanifold is governed by its {\em mean curvature vector} field, and thus trapped submanifolds can be characterized by properties of this vector field. 

To be more precise, consider any smooth $(n-m)$-dimensional submanifold $\zeta\subset M$ embedded in the $n$-dimensional spacetime $M$. Let $\{\vec e_A\}$ be a basis of vector fields tangent to $\zeta$ and denote by $\gamma_{AB} = g\vert_{\zeta} (\vec e_A, \vec e_B)$ the (components in this basis of the) first fundamental form of $\zeta$. We assume that $\zeta$ is spacelike, so that $\gamma_{AB}$ is positive definite and $(\zeta, \gamma)$ is a Riemannian manifold on its own whose Levi-Civita connection will be denoted by $\overline\nabla$, while the connection in $(M,g)$ is $\nabla$ as usual. Therefore, by splitting into tangential and normal parts with respect to $\zeta$ we have \cite{Kri,O,S2,SMilestone}
$$
\nabla_{\vec e_A} \vec e_B = \overline\nabla_{\vec e_A} \vec e_B - \vec K_{AB} 
$$
where $\vec K_{AB}$ is called the shape tensor, or the second fundamental form vector, of $\zeta$ in $M$. Notice that $\vec K_{AB}$ is normal to $\zeta$. Given any one-form $\bm{v}$ defined at least on $\zeta$ one also has on $\zeta$
\be\label{nablas}
e^\mu_A e^\nu_B \nabla_\mu v_\nu = \overline\nabla_A \bar{v}_B +v_\mu K^\mu_{AB} 
\ee
where $\bar{\bm{v}}$ denotes the pull-back of $\bm{v}$: $\bar v_A := \bm{v} (\vec e_A)$. 

The mean curvature vector of $\zeta$ in $M$ is the trace of $\vec K_{AB}$, that is
$$
\vec H := \gamma^{AB} \vec K_{AB} .
$$
Observe that $\vec H$ is normal to $\zeta$ by construction, and therefore it has $m$ independent components. If $\vec n$ is any vector field normal to $\zeta$
$$
\theta_n := n_\mu H^\mu 
$$
is called the {\em expansion} along $\vec n$ of $\zeta$. In particular if $\vec n$ is null, $\theta_n$ is called a {\em null expansion}. Contracting (\ref{nablas}) with $\gamma^{AB}$ a formula using the Lie derivative of the metric can be derived
\be\label{nablas2}
\frac{1}{2} \Pi^{\mu\nu} (\pounds_v g)_{\mu\nu} = \overline\nabla_A \bar{v}^A +v_\mu H^\mu 
\ee
where
\be\label{Pi}
\Pi^{\mu\nu} := \gamma^{AB} e^\mu_A e^\nu_B 
\ee
is the (contravariant) projector to $\zeta$.

The mean curvature vector controls the variation of the $m$-volume $V_\zeta$ of $\zeta$ along any possible deformation vector. Let $\vec \xi $ be any vector field and deform $\zeta$ along its flow. Then, on using (\ref{nablas2}), the initial variation of the $m$-volume can be easily seen to be \cite{Kri,O}
$$
\delta_\xi V_\zeta = \int_\zeta \left(\overline\nabla_A \bar\xi^A + \xi_\mu H^\mu \right)
$$
so that for closed (compact) $\zeta$ the first term vanishes and we are left with
\be\label{varvol}
\delta_\xi V_\zeta = \int_\zeta \xi_\mu H^\mu 
\hspace{1cm} \mbox{for compact $\zeta$}.
\ee
The classical characterization of {\em minimal} (or rather {\em extremal}) submanifolds by the vanishing of $\vec H$ follows automatically. In Riemannian geometry, this is the only distinguished case. However, in Lorentzian geometry, $\vec H$ can also be timelike, or null, and this provides new important types of submanifolds. This leads to the definition of trapped submanifolds, and their avatars.

A (closed) submanifold $\zeta$ of any dimension is said to be \cite{S2,SMilestone}
\begin{itemize}
\item future trapped if $\vec H$ is future timelike everywhere on $\zeta$,
\item weakly future trapped if $\vec H$ is future causal everywhere on $\zeta$,
\item marginally future trapped if $\vec H\neq 0$ is future null everywhere on $\zeta$.
\end{itemize}
and similarly for past trapped. Notice that, from (\ref{varvol}), any closed future trapped submanifold has a negative variation of $m$-volume along {\em any possible future direction $\vec \xi$}. And that {\em all future null expansions} are negative. This connects with the original definition by Penrose, who stated that a trapped (co-dimension 2) surface is trapped if two independent null expansions are negative\footnote{The original definition is still used in most of the physics literature, and in a large part of the mathematical one. The reasons behind this are obscure to me. The characterization with the mean curvature vector is clearly neater and provides more information, apart from unifying the concept for arbitrary dimensional manifolds. A (future) trapped submanifold has \underline{all} possible (future) expansions negative, not only the null expansions. Moreover, the computation of null expansions and the choice of null directions complicates the explicit calculations. Computing the mean curvature vector is far simpler! \cite{S2002}}.

The characterization with the mean curvature vector has many advantages. As an example, one can immediately derive that there cannot be any closed (weakly, marginally) trapped surfaces contained in stationary regions of the spacetime \cite{MS}. For, by taking $\vec \xi$ in (\ref{varvol}) as the timelike Killing vector, we know that $\delta_\xi V_\zeta =0$, and thus $H^\mu$ cannot be future, nor past, everywhere on $\zeta$ (as if it were, $\xi_\mu H^\mu$ would have a sign). Similar results can be obtained if there are null Killing vectors, or causal conformal Killing vectors, etc. \cite{BeS,MS,S2003}. As an illustrative example, and as mentioned previously, one can easily prove that plane waves (\ref{PW}) cannot contain any closed trapped submanifolds. This follows because the null vector field $\vec k$ is parallel $\nabla_\nu k^\mu=0$, hence in particular is a hypersurface-orthogonal Killing vector, so that using $\vec\xi =\vec k$ in (\ref{varvol}) one again derives that the integral on $\zeta$ of $k_\mu H^\mu $ vanishes and therefore $H^\mu$ cannot be timelike everywhere. Many other interesting results using (\ref{nablas}) and the mean curvature vector can be found, see \cite{BeS,MS,S2}.

Now that we know that the concept of trapped submanifold is independent of the co-dimension, we may come back to the question raised above in subsection \ref{subsec:dim}: why should there only be a few co-dimensions in the boundary assumption of the incompleteness theorems?

\subsection{The curvature condition. ``Energy'' conditions}\label{subsec:EnergyCond}
Let me finally analyze hypothesis 4, the {\em convergence condition}. This is an assumption on the curvature tensor which takes care of the attractive character of the gravitational field. Traditionally, the curvature condition has been extracted from the Raychaudhuri equation \cite{Ray,Ray2,K,HE,S1} as a requirement to ensure the so-called focusing effect of gravity for generic congruences of causal geodesics. If these congruences of causal geodesics emanate from a point, or orthogonally from a co-dimension one (for timelike geodesics) or co-dimension 2 (for null geodesics) submanifold, the curvature condition is simply (\ref{ConCon}) where $v^\mu$ is the vector field tangent to the geodesic congruence. If this holds, and if the initial expansion --or divergence/convergence--- of the geodesic congruence is negative (respectively positive), a {\em caustic} develops in finite proper time, or affine parameter for null $v^\mu$,  to the future (resp.\ past). These are usually called {\em conjugate and  focal points}. An important consequence of this is that causal geodesic curves, and causal geodesic congruences orthogonal to (hyper)-surfaces, stop being maximizing if they encounter one of these caustics \cite{BEE,HE,P5}. One can prove that, for a closed future-trapped surface, this implies that the boundary of the future\footnote{Here, what I mean by the `boundary of the future' of $\zeta$ is $E^+(\zeta)$ \cite{HE,HP,P5,S1}, defined as the set of points that can be reached from $\zeta$ causally, but not through a timelike curve. If $E^+(\zeta)$ is compact, then $\zeta$ is a {\em trapped set}. Not to be confused with a trapped submanifold. Trapped submanifolds become trapped sets precisely {\em under the appropriate curvature condition} --if spacetime is null complete--, as explained in the text.} of the surface is compact (if the spacetime is complete). Such surfaces are thus examples of {\em trapped sets}, a key concept introduced in the powerful Hawking-Penrose theorem \cite{HP,HE}. 

Nevertheless, the points focal to a submanifold $\zeta$ are best controlled (or identified) by using the {\em index form} \cite{Kri,O,GaS}, and this is actually {\em completely independent of the co-dimension} of $\zeta$. In particular, one can prove (proposition 1 in \cite{GaS}) that there is a point focal to $\zeta$ along any future-directed geodesic emanating orthogonally from $\zeta$ with tangent vector $N^\mu$ and initial negative expansion if 
\be\label{cond}
R_{\mu\nu\rho\sigma}P^{\nu\sigma} N^\mu N^\rho \geq 0
\ee
(provided that the geodesic gets that far) where $P^{\nu\sigma}$ is the parallel propagation of the projector $\Pi^{\nu\sigma}$ defined in (\ref{Pi}) along the geodesic:
$$
N^\rho\nabla_\rho P^{\mu\nu} =0, \hspace{1cm} P^{\mu\nu}\vert_\zeta = \Pi^{\mu\nu} .
$$
This readily implies that, if (\ref{cond}) holds along the future-directed null geodesics starting orthogonal from a compact $\zeta$, then $\zeta$ is a (future) trapped set (proposition 3 in \cite{GaS}) if $\zeta$ is a trapped submanifold, {\em independently of the dimension of $\zeta$} ---or there are incomplete null geodesics.

As the existence of a trapped set is the basic ingredient in the proof of the Penrose and the Hawking-Penrose theorems \cite{HP,HE,S1}, it follows that actually one can assume the existence of a trapped submanifold of any dimension together with the curvature condition (\ref{cond}) to get the same incompleteness result. Thus, the standard singularity theorems were generalized to the case of having a trapped submanifold of any co-dimension in \cite{GaS}. Later, these generalized theorems were also proven in low differentiability \cite{GGKS,KOSS}.

Some relevant remarks are needed to clarify the role of condition (\ref{cond}). First of all, from a physical point of view (\ref{cond}) is a condition on tidal forces (mathematically speaking, this a condition on sectional curvatures) \cite{GaS}. If (\ref{cond}) holds, the tidal forces, alternatively the geodesic deviations, in directions (initially) tangent to $\zeta$ are attractive on average. The overall result is a tendency to converge that represents the attractive property of gravity.

Secondly, one must notice that, for co-dimension one, that is to say a hypersurface $\zeta$, there is a unique (timelike) normal direction $N^\mu$  and therefore $P^{\mu\nu}=g^{\mu\nu} -N^\mu N^\nu /(N^\rho N_\rho)$ so that (\ref{cond}) becomes simply $R_{\mu\nu} N^\mu N^\nu \geq 0$, the convergence condition (\ref{ConCon}) along $N^\mu$. Exactly the same happens with surfaces, co-dimension two $\zeta$ \cite{GaS}. One thus recovers the classical theorems automatically when $\zeta$ has co-dimension one or two. 

The generalization to trapped submanifolds of any co-dimension is relevant to yet another of Penrose's ideas \cite{Pextra}: the possible classical  instability of compact extra dimensions ---such as those necessary for consistency in string theory. According to Penrose, such compact extra dimensions might develop `singularities', in the sense of geodesic incompleteness, in extremely short times. Similar arguments were raised in  \cite{CGHW} for the case of large, even infinite, extra dimensions, and there is also the pionnering result in \cite{Wit} for semi-classical perturbations. Penrose's argument was appealing, but somehow it needed some extra assumptions, such as unnatural splittings of the spacetime in order to be able to connect with the Hawking-Penrose theorem. The main problem here was that one needs co-dimension one, two or $n$ (a point) to apply the theorem, but of course the extra-dimensional space has co-dimension four (and submanifolds within the extra-dimensional space a larger one). This is where the generalized theorems are relevant, as they admit all possible values of the co-dimension. Such a possibility was suggested in \cite{GaS} and later thoroughly analyzed in \cite{CiS} , where incompleteness theorems for warped products were proven and studied in connection with the instability of product manifolds. The idea was to check if a dynamical evolution of the compact extra-dimensional part, considered as a perturbation of a direct product spacetime, would lead to singularities. The answer is yes under some precise circumstances, but not in full generality \cite{CiS}. These kind of `instabilities' are of a different kind as those arising from initial value formulations and the analysis of how perturbations may grow, as there are important results showing the (classical) stability of product spacetimes in that sense, such as \cite{Wy,BFK} and the more recent \cite{ABWY} for a high ($n\geq 11$) dimensional direct product of $n\geq 10$ Minkowski spacetime times a compact Ricci-flat Riemannian manifold --if the entire perturbed spacetime is also Ricci flat. In any case, this is yet another example of the influence that Penrose's arguments have had in the general gravitational community.

A final general comment about the curvature condition. In the traditional formulations of the theorems, and in most of the standard literature, this assumption is termed an {\em energy condition}. The reason is clear when one is working in GR, for in GR there is the direct relation between the Ricci tensor and the energy-momentum tensor given by the field equations (\ref{eq:EFE}). Thereby, the convergence condition (\ref{ConCon}) can be rewritten in terms of physical quantities related to $T_{\mu\nu}$. However, the curvature condition is a restriction on the curvature tensor (or its trace, the Ricci tensor), which is a geometrical object. Hence, the theorems are valid in general geometric theories of gravity based on Lorentzian geometry, or even generalizations thereof such as Finsler-like geometries, see e.g. \cite{AJ}.

 \subsection{A 2022 version of Penrose's theorem}
 
 A modern version of the 1965 singularity theorem \cite{P}, incorporating many of the different improvements mentioned so far that have allowed for a better understanding of the underlying mathematical and physical ideas, could be as follows
\begin{theorem}[Sketch of a `2022 theorem']\label{th:modern}
Let $(M,g)$ be an $n$-dimensional Lorentzian manifold with a metric $g$ of {\color{blue} class $C^1$}. If
\begin{enumerate}
\item there is a $C^2$ non-compact Cauchy hypersurface ({\color{blue} alternatively, spacetime is past-reflecting and spatially open})
\item there exists a future-trapped $C^2$ submanifold $\zeta$ of {\color{blue} any dimension}
\item the {\color{blue} tidal condition (\ref{cond})} holds along future-directed null geodesics emanating orthogonally from $\zeta$
\end{enumerate}
Then, there are future-incomplete null geodesics.
\end{theorem}
\noindent\underline{Warning:} As stated, this is just a sketch of the actual theorem, highlighting the novelties, because the technical details behind every condition are cumbersome. For instance, the curvature tensor is not a tensor field but rather a tensor distribution, so that condition (\ref{cond}) cannot hold pointwise: it has to be appropriately generalized by using regularizations. Similarly, geodesics (as solutions of an ODE system) are no longer unique, and focal points are not well defined, just to mention a few problematic steps. However all these difficulties can be resolved satisfactorily and, more importantly, the physical meaning of all the hypotheses and the conclusion remain the same. For the technical details, I refer to \cite{GaS,Graf,KOSS,stein}. Similarly, the theorem using the version with past reflectivity and open space was only proven for co-dimension 2 trapped submanifold \cite{Mi} under the null energy condition, but it may well hold in for general trapped submanifolds with condition (\ref{cond}).

\section{Assessment of the 1965 theorem: merits and misconceptions}
In my opinion, the greatest merit of the 1965 theorem relies on the introduction and use of the concept of closed trapped surface. It has been a prolific idea with multiple applications, see e.g. \cite{Bengtsson,Dafermos,HayBook,JVG,S2}, as well as an object of interest in pure mathematics \cite{eichmair}.
It has evolved into a large fauna of interesting surfaces that generalize the minimal surfaces of classical geometry \cite{AMS1,AMMS,Ne3,S0}, and also led to the local concepts of isolated, dynamical, trapping,... horizons \cite{AK,AMS,AG,Booth,BeS,Hay,Jara}. Thereby, it is also used to detect the formation of black holes in {\em Numerical Relativity} \cite{BSh,JVG,Thor} or changes in
their evolving horizons. Similarly, it has influenced the development of gravity analogues \cite{BLV} and, of course, 
it also plays a role in the analysis of the Cosmic Censor Conjecture \cite{Wald1,landsman}. In relation to this, it fostered many important geometric and physical inequalities, such as for instance the Penrose inequality \cite{PI,Mars}, but many other `isoperimetric-like' inequalities too \cite{CLW,Dain,Dain1,Dain2,DJR,DaRe,GJR,Gi,Gi1,JRD,Malec,Simon}.

As impressive as this list may seem, this is not the full story. There is an even more important point concerning trapped submanifolds which is key to the theorem and its implications: {\em stability}.

The notion of trapped submanifold is intrinsic, a pure geometric notion, and its characterization is the {\em inequality} $H^\mu H_\mu<0$.\footnote{If one prefers the traditional notion with negative expansions, they are also strict inequalities} Hence, trapped submanifolds are {\em stable} under small perturbations of the spacetime. This was already emphasized by Penrose himself, by comparing the {\em stability} of censored versus naked singularities \cite{P4}:
\begin{quote}
However, there is an essential difference between the logical status of the [naked] 
singularity (...) and that [censored (...)]. In the [censored] cases (...)  there are trapped surfaces present, so
we have a theorem which tells us that even with generic perturbation a singularity
will still exist. In the [naked] situation (...), however, we have no trapped surfaces,
no known theorem guaranteeing singularities (...)
\end{quote}
Or again in (\cite{PI}, p.133):
\begin{quote}
It might even be possible to produce exact solutions representing a collapse of some special
matter distribution to such a [naked] singularity. But this would in itself tell us rather little.
We would have to know whether such behavior was ``stable'' in the sense of being
qualitatively unchanged when the initial conditions are perturbed in some small but
finite way.
\end{quote}

\begin{figure}[!h]
\centering
\includegraphics[width=5.9cm]{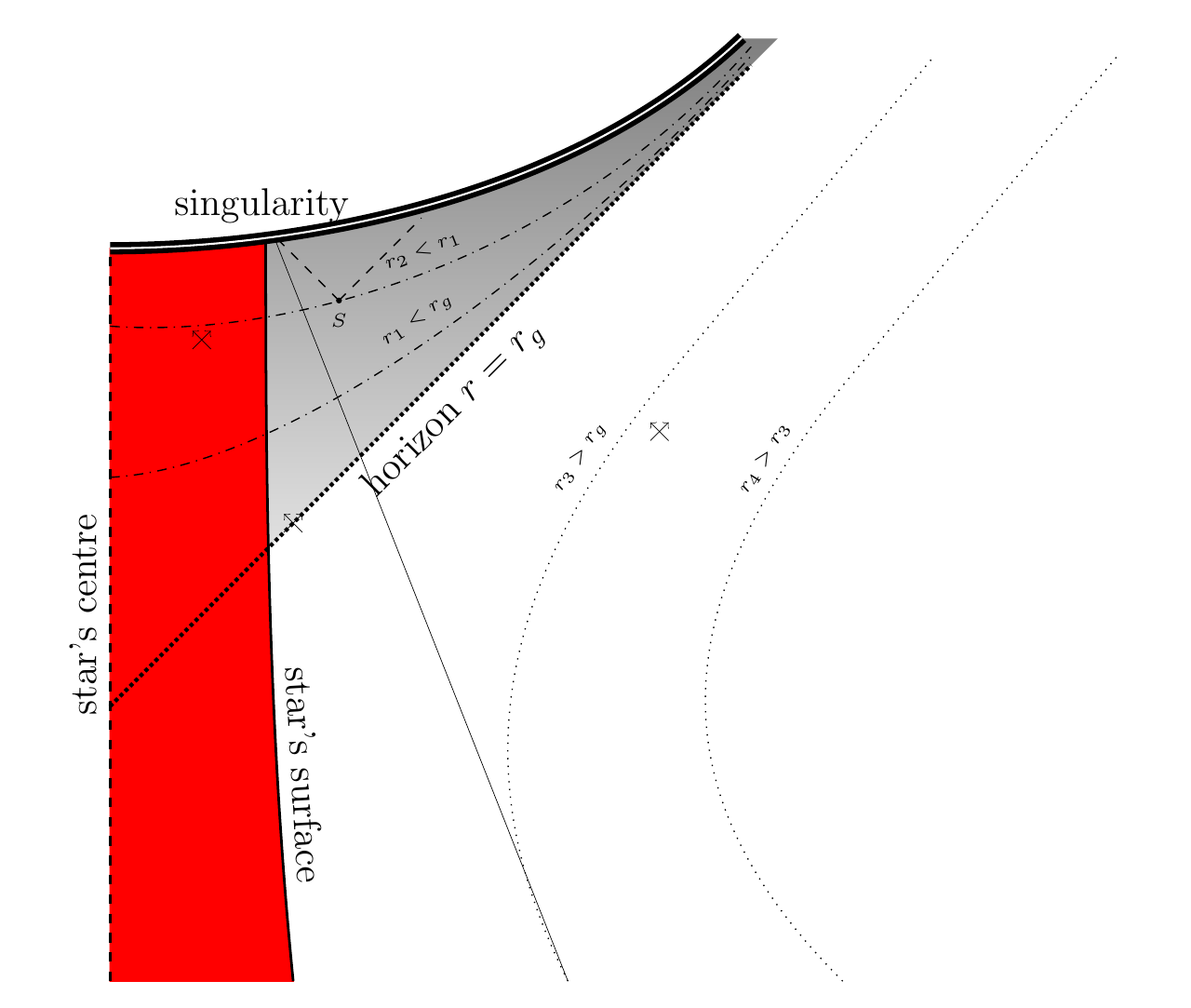}
\includegraphics[width=5.9cm]{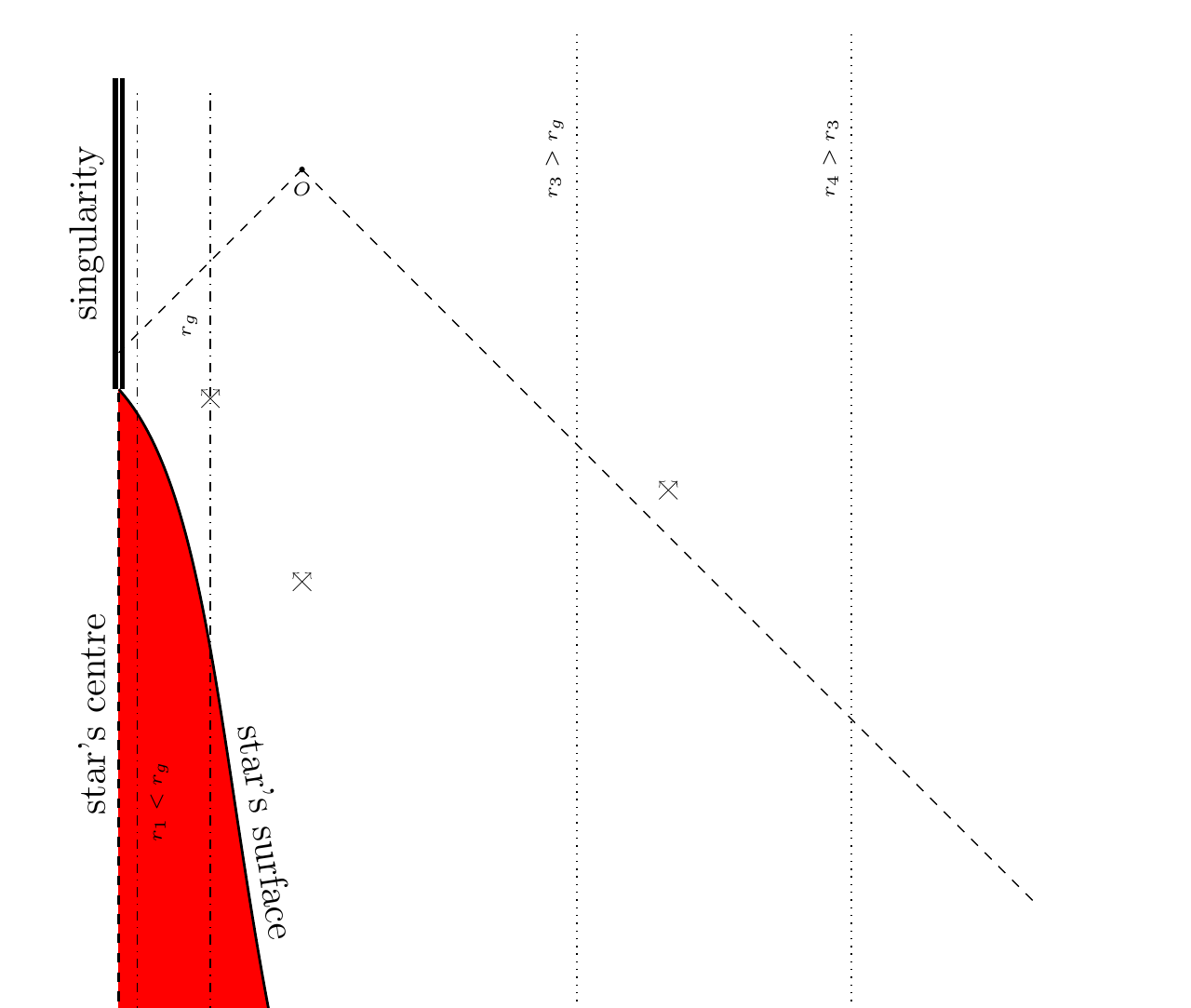}
\caption{(Almost) spherically symmetric collapse of matter (in red) to form a black hole (left) or a naked singularity (right). In the black-hole spherical case, the exterior field is (\ref{Schw}). On the right, one may think, for the purposes of illustration, that the matter collapsing has total negative mass. Only the radial part is shown, as usual, and time increases upwards. The round spheres have area $4\pi r^2$ and hypersurfaces with constant $r$ are depicted in both cases. Light cones are at 45$^o$ as shown with the arrowed little crosses. In the standard, censored, case on the left there are trapped spheres, such as $S$ shown, and their stability plus the singularity theorem imply that the future singularity is also {\em stable} against small perturbations. Notice that this singularity is not visible outside the horizon. Nothing of the kind happens on the right, which contains no closed trapped surfaces and the singularity is visible, for instance for the observer $O$. The stability of such a  situation is doubtful.}\label{fig1}
\end{figure}

That the stability of trapped submanifolds is a key point can be better understood by noticing that there are well-known models of realistic {\em spherically symmetric} collapse to produce black holes --see \cite{HE,Wald,H4} for the definition of a black hole-- with a singularity in the future, the paradigmatic example is the Oppenheimer and Snyder \cite{OS,OV} dust model, but there are many others with (perfect) fluids, e.g. \cite{FST}.  Of course, the assumption of spherical symmetry raises some reasonable doubts about how robust and generic these results might be. With Penrose's theorem this is clearly resolved: the mentioned spherical models contain closed trapped surfaces, and the Einstein-Euler set of field equations describing a perfect fluid in GR constitutes a system of hyperbolic PDEs, so that the property of continuous dependence of the solution on the initial conditions holds.  Given that the initial conditions of the spherically symmetric models lead to a trapped sphere in finite proper  time, it follows that initial conditions sufficiently close to the spherical ones also give rise to closed trapped surfaces that must apear within the same time interval, {\em regardless of symmetries}. This is illustrated in Figure \ref{fig1}. Of course, this applies to situations of sufficient differentiability, so that the improvements lowering differentiability of the theorems discussed above might not lead to such stability outcomes.

Now let me consider some of the misconceptions and folklore surrounding Penrose's, and the general, singularity theorems. First of all, the theorems are many times considered to prove that black holes form when gravitational collapse occurs with high remaining mass, this is especially so in the astrophysics community. This is far from the truth, despite sentences such as the statement for which Penrose was awarded half of the Nobel prize in Physics 2020: ``{\em for the discovery that black hole formation is a robust prediction of the general theory of relativity}'', see also the discussion in \cite{landsman}. This may be argued to be right if all contributions by Penrose to GR --including the yet unproven cosmic censorship conjecture-- are taken into account but, as I just wrote, this is incorrect in various respects if it refers to his singularity theorem exclusively. Firstly, and most importantly, because of the assumption of the existence of a closed trapped surface. It actually happens that, in asymptotically flat situations, closed trapped surfaces are completely enclosed beyond the event horizon, which is the distinctive feature of a black hole. There is actually a theorem \cite{HE,C,Wald} proving this under some reasonable circumstances. In simpler words, closed trapped surfaces lie entirely within the black hole region of a black hole {\em already formed}. Therefore, the theorems are speaking about the {\em interior of black holes}, or about spacetimes that are sure to contain a black hole.

Of course, Penrose was aware of all this, see e.g. the following passage taken from \cite{P7}, pp.233-4 (italics are mine)
\begin{quote}
It is not hard to conceive of physical situations in which one of the standard criteria
for `unstoppable gravitations [sic] collapse' is satisfied. All that is required is for sufficient
mass to fall into a small enough region. For the central region of a large galaxy, for
example, the required concentration could occur with the stars in the region still
being separated from each other, so there is no reason to expect that there could be
some overriding physical principle which conspires always to prevent such unstoppable collapse. {\em However, we cannot simply deduce from this that a black hole will be the result}. This deduction requires the crucial assumption that cosmic censorship,
in some form, holds true.

(... ...)

It appears to be a not uncommon impression among workers in the field that as
soon as one of these conditions is satisfied ---say the existence of a trapped surface--- then
a black hole will occur; and, conversely, that a naked singularity will be the
result if not. However, it should be made clear that neither of these deductions is in
fact valid. The deduction that a black hole comes about whenever a trapped surface is
formed requires the assumption of cosmic censorship.
\end{quote}

Therefore, the important question to be answered is whether or not closed trapped surfaces form in the evolution of stable and completely innocuous, regular initial data. This was addressed in the massive work by Christodoulou \cite{Chris1}, see also \cite{RT,KR}.

Secondly, even more radical statements are too many times read or heard, such as ``singularities are consubstantial to GR'' or similar ones, and these blunt statements come inevitably accompanied by a reference to the singularity theorems. In this sense, it is worth recalling that (almost) all gravitational systems, say planets, stars, planetary systems, constellations, clusters, galaxies, pulsars, binary systems, ...  are non-singular. They are correctly described by GR and its post-Newtonian or post-Minskowskian limits. In fact, (post)-Newtonian gravity is free from the singularity problem. What the singularity theorems (in GR and any other gravitational theory based on a Lorentizian manifold) have surfaced is the existence of two feasible situations where something strange, or not understood, happens: the extremely early Universe --if the Universe is on average as we nowadays think it is and it has been-- and the mysterious, because of their inaccessibility, black holes' deep interiors. The incompleteness predicted by the theorems in these two extreme situations is certainly something found out by means of the theorems.

And once more, this was clear to Penrose. I quote now from \cite{P4}
\begin{quote}
The main significance of [the singularity] theorems (...), is that they show that
the presence of space-time singularities in exact models is not just a feature of
their high symmetry, but can be expected also in generically perturbed models.
This is not to say that all general-relativistic curved space-times are singular 
---far from it. There are many exact models known which are complete and free
from singularity. But those which resemble the standard Friedmann models or the
Schwarzschild collapse model sufficiently closely must be expected to be singular.
\end{quote}

To end this section, I want to stress a couple of shortcomings in the theorems, which are related to each other: the very mild conclusion, and the unsolved question of spacetime extendibility. 

Concerning the former, and against popular belief, there is no definite established relationship between
incomplete geodesics and curvature divergences. Sometimes, limits on curvature growth can be placed on
{\em maximal} incomplete geodesics \cite{KRa,Ne,Szab,T2}. Concerning curvature divergences there are scarce results \cite{Cla4,CS,Tho,TCE}. However, in many cases, what the theorems predict are some kind of horizon (usually a Cauchy horizon) which can be made perfectly regular by extending the spacetime beyond. In other words, it might be the case that the manifold itself is incomplete, because one has inadvertently cut out regular portions thereof.

Which brings me to the second point, the important physical problem of how to handle extendible space-times. This is a complicated matter \cite{SMilestone,S1,S6}. The main difficulty is that the possible extensions are highly non-unique. As an illustrative example, consider the exterior part of the Schwarzschild solution (\ref{Schw}). We are all used to see the Kruskal-Szekeres {\em maximal analytical} extension in textbooks, e.g. \cite{MTW,Wald,HE}, by keeping spherical symmetry and a vanishing Ricci tensor. But, does such an extension makes physical sense? What one obtains is a globally hyperbolic, {\em completely empty }spacetime, with two asymptotic regions, a strange topology and two curvature singularities, one in the past the other in the future. Not very realistic! Many other extensions are actually available, for instance at the end of \cite{S1} one could count at least eleven different extensions of (\ref{Schw}). Herein, let me just show an example of extension leading to what is usually called a spherically symmetric regular black hole. There are many more such regular black holes, see e.g. \cite{Maeda} and references therein, but as far as I know this was the first example keeping the entire exterior region $r>r_g$ of (\ref{Schw}) and at the same time keeping the weak energy condition, namely
$$
(R_{\mu\nu} -\frac{1}{2} R g_{\mu\nu}) v^\mu v^\nu \geq 0
$$
for any causal $v^\mu$. Notice that, for null $v^\mu$ this is simply the null convergence condition (\ref{ConCon}). The model is given in standard advanced coordinates by  \cite{MMS}
\be\label{eq:BHregular}
ds^2= -e^{\beta(r)} \left(1-\frac{2m(r)}{r}\right) dv^2 +2dvdr +r^2 \left(d\theta^2 +\sin^2 \theta d\varphi^2 \right)
\ee
where the functions $\beta$ and $m$ read explicitly
\bean
2m(r) &=& r_g \Theta(r-r_g) + \frac{r^3}{r_g^2} \left(10-15\frac{r}{r_g} +6\frac{r^2}{r_g^2} \right)\Theta(r_g-r),\\
\beta (r) &=& \frac{5}{3}\left(1+3\frac{r}{r_g}\right)\left(\frac{r^3}{r_g^3} -1\right)\Theta(r_g -r) 
\eean
where $\Theta (x)$ is the step Heaviside function. This spacetime is properly matched across the null hypersurface $r=r_g$ so that the metric is $C^{1,1}$ \cite{MS0}. Smoother models could also be built. Note that the metric is exactly (\ref{Schw}) for all $r\geq r_g$ --in advanced coordinates--- having $\beta =0$ and a constant mass function $2m = r_g$ there. 

\begin{figure}[!h]
\centering\includegraphics[width=8cm]{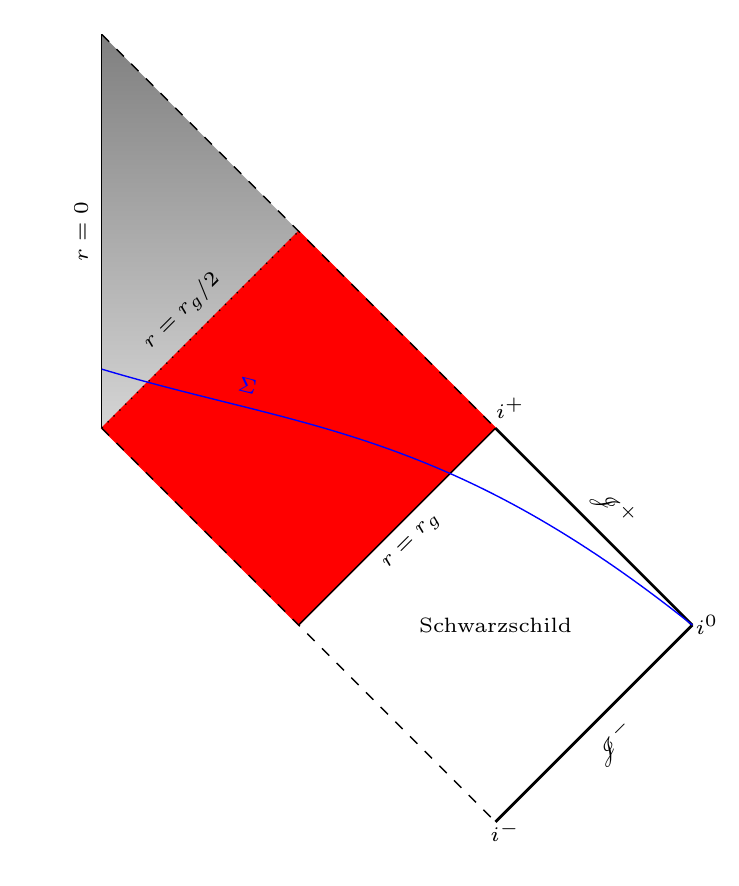}
\caption{A conformal diagram of the spacetime (\ref{eq:BHregular}). The line $r=0$ is a regular centre. The non-colored region represents exactly (\ref{Schw}). The red region contains trapped round spheres. The hypersurface $\Sigma$ is a non-compact Cauchy hypersurface. Causal geodesics reaching the dashed lines, to the future and to the past, are incomplete.}
\label{fig:BHregular}
\end{figure}

The extended portion of the spacetime is the region $r< r_g$, see figure \ref{fig:BHregular}, where the Ricci tensor no longer vanishes. As already mentioned, the {\em null} convergence condition (\ref{ConCon}) is satisfied. There are trapped round spheres whenever $r_g/2 < r < r_g$, and the curvature tensor is regular everywhere with a regular centre of symmetry at  $r=0$. This spacetime also possesses non-compact Cauchy hypersurfaces, as the $\Sigma$  shown in figure \ref{fig:BHregular}. In summary, all the hypotheses of the singularity theorem are met, ergo some null geodesics must be future incomplete.  This is the case for instance for many radial null geodesics other than $v=$ const., as they reach $r=r_g/2$ for $v\rightarrow \infty$ with finite affine parameter. There are also past incomplete null geodesics, but these are not predicted by the theorem. This spacetime can be further regularly extended in many ways, and geodesically-complete {\em extensions} exist, see \cite{MMS,S6}. These larger complete extensions avoid the Penrose singularity theorem because the non-compact $\Sigma$ becomes a {\em partial} Cauchy hypersurface. The more general theorems \cite{HP} are not applicable either because (\ref{ConCon}) does not hold for arbitrary causal $v^\mu$.

\begin{figure}[!h]
\centering\includegraphics[width=12cm]{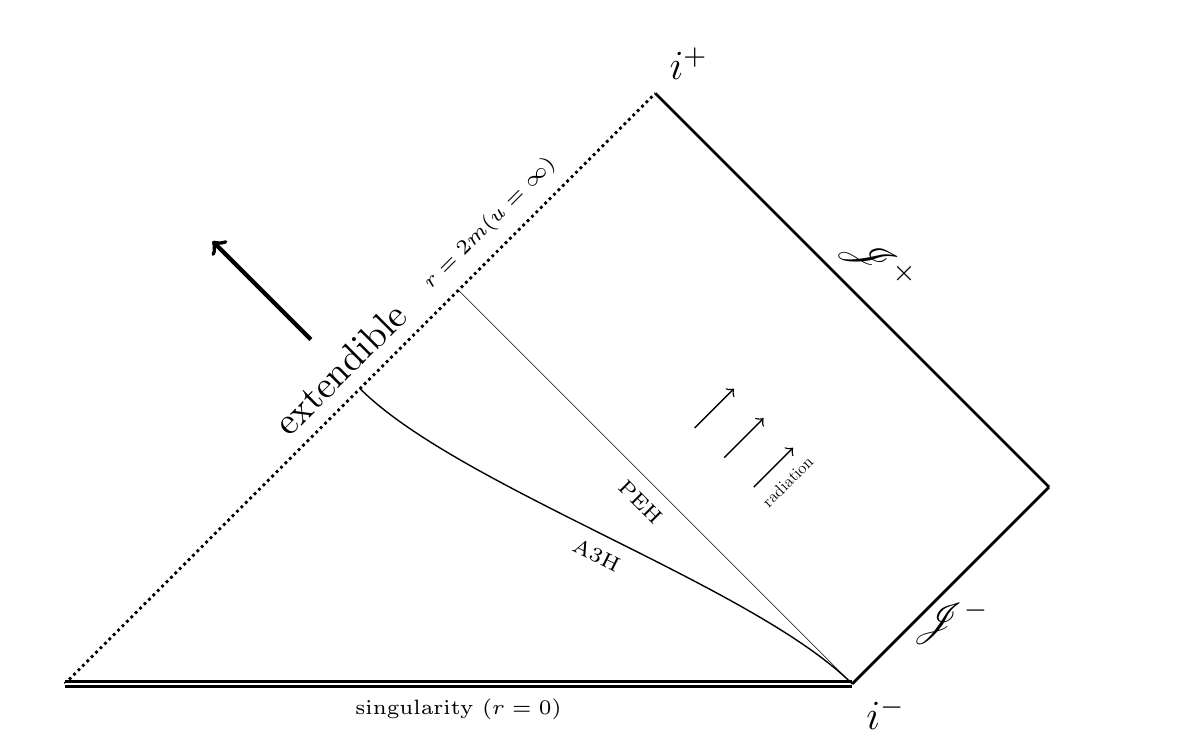}
\caption{A conformal diagram of the spacetime (\ref{Vaidya}). There is a singularity with $r=0$ in the past. There is a dynamical (or apparent) horizon A3H at $r=2m(u)$, as shown, which diverts from the past event horizon PEH. Radiation is emitted along $u=$const.\ lines so that the mass function $m(u)$ is a decreasing function of $u$.  Causal geodesics reaching the dotted line in the future, which has an areolar coordinate $r=2m(u\rightarrow \infty)$, are incomplete. The metric can be extended beyond this line, but how?}
\label{Vaidya1}
\end{figure}

As explained, the metric (\ref{eq:BHregular}) is just one possible extension among infinitely many that can be attached to the original Schwarzschild metric. Physically meaningful extensions are extremely difficult to identify, and it also may depend on personal taste. Should we always prefer extensions that keep classical energy conditions and lead to singularities, over other extensions that avoid the singularities at the price of violating some classically reasonable conditions? From the mathematical point of view, analytical extensions might be an answer, but they are not always available, nor they are unique in general. As another illustrative example, consider the exterior radiating Vaidya metric given in standard retarded coordinates by \cite{Exact}
\be\label{Vaidya}
ds^2= - \left(1-\frac{2m(u)}{r}\right) du^2 -2dudr +r^2 \left(d\theta^2 +\sin^2 \theta d\varphi^2 \right)
\ee
where $m(u)$ is a mass function depending on retarded time $u$. This spacetime satisfies (\ref{ConCon}) as long as $m(u)>0$ is a non-increasing function of $u$. If $m$ were constant this would simply be Schwarzschild. If there is a lower  non-vanishing bound $\mu >0$ for $m(u)$, that is to say, $\mu:=\lim_{u\rightarrow \infty} m(u)$, one can easily check that timelike radial geodesics reach $r=2\mu$ in finite proper time, so that they are incomplete, see figure \ref{Vaidya1}. 

The metric is thus extendible beyond that region $r=2\mu$, but the extension to be performed is far from obvious and one needs some physical information to perform it. A detailed discussion can be found in \cite{FMS}. This problem was first studied by Israel \cite{Israel}, who pointed out that his approach to the problem was not fully
satisfactory because the extended spacetime was chosen beforehand. However, observers in the metric (\ref{Vaidya}) can only get to know the properties of $m(u)$ for finite values of $u$, while to build the extension they should somehow guess at least some properties of the mass beyond $u\rightarrow \infty$. Furthermore, even knowing $m(u)$ for all possible values of $u$ plus some of the physical properties of the particular situation to be described is not enough to define a unique extension. As proven in \cite{FMS}, by keeping the spherical symmetry and a null radiation energy-momentum tensor, the ambiguities in the extension amount to the choice of the mass function beyond $r=2\mu$, which then becomes a future event horizon. In general, there will exist many different choices for the mass beyond that horizon that comply with all the physical requirements. And, by the way, analytical extensions are many times not even possible. 

\begin{figure}[!h]
\centering\includegraphics[width=6cm]{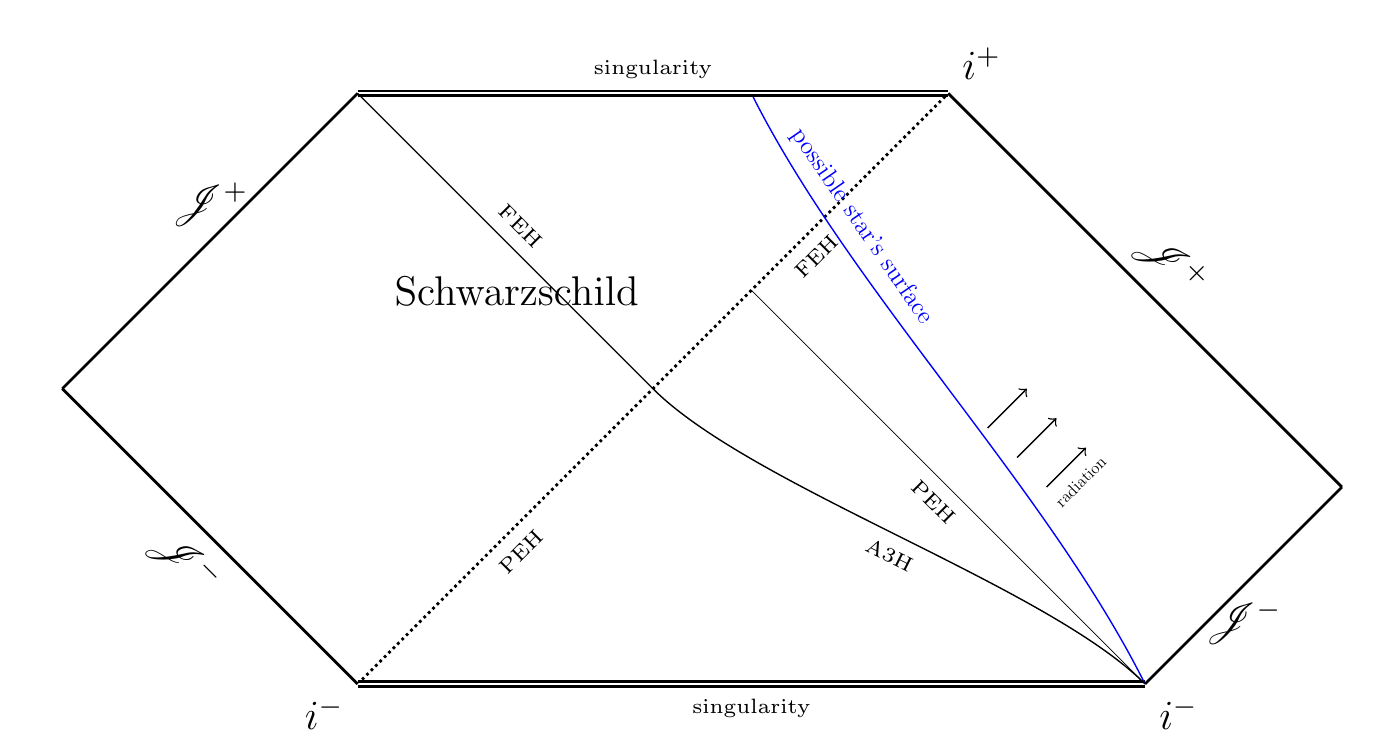}\includegraphics[width=6cm]{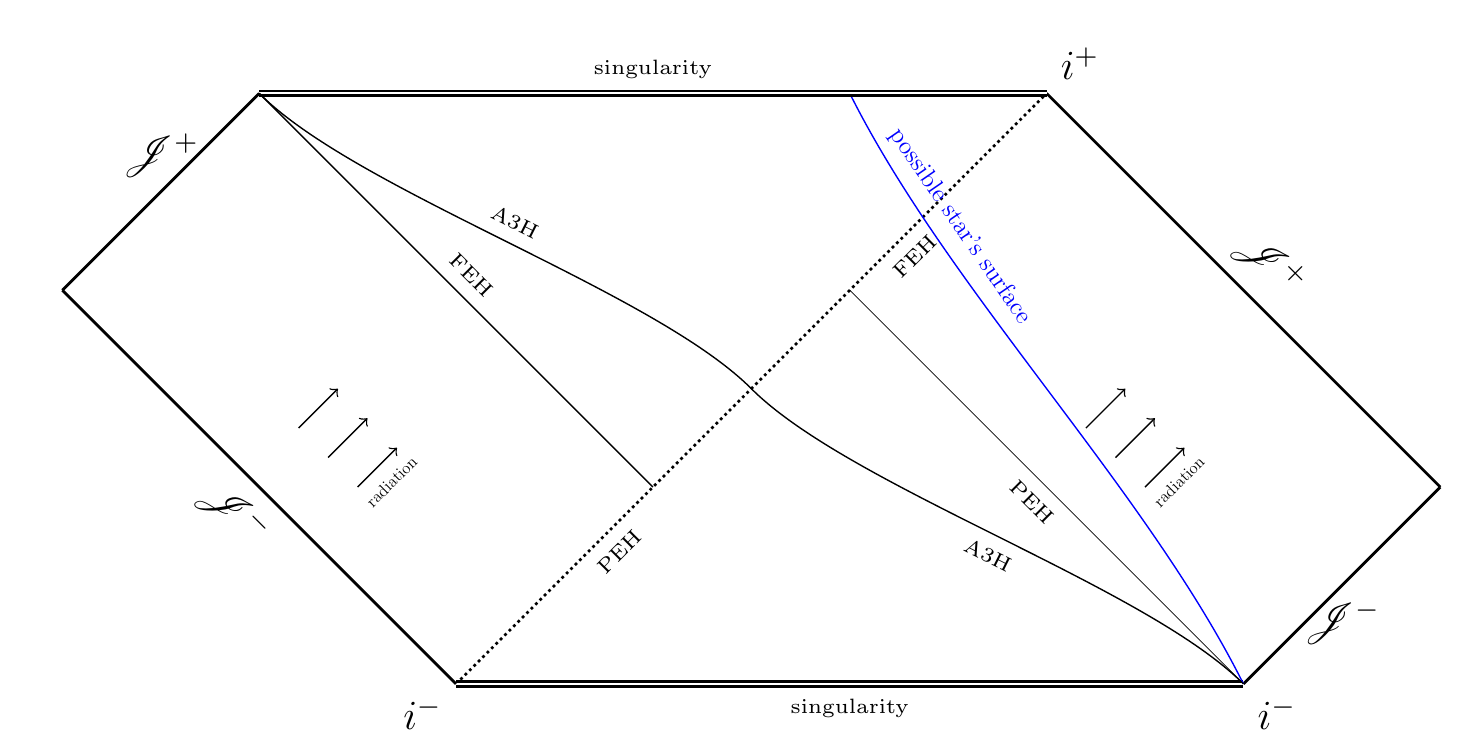}
\caption{Two posible extensions, among many, of the metric (\ref{Vaidya}) of figure \ref{Vaidya1}. The hypersurface through which the extension is performed becomes a partly future, partly past, event horizon FEH/PEH in both cases, and there arises a second singularity with $r=0$ in the future. On the left the mass function is kept constant at equal to $\mu$ all along, so that the added part of the spacetime is actually Schwarzschild, the original A3H merges with a new FEH and there is no radiation there. On the right an alternative completely different situation is represented. Now the mass function has a minimum at $r=2\mu$ and the added portion of the spacetime is mirror symmetric with the original one. Thus, radiation keeps flowing though, in the extended part, it now has an `incoming' character. In this case, the A3H is a dynamical horizon everywhere except at the central point where it becomes null instantaneously. Unless one has some information about the processes behind the radiation, it is not possible to choose between these two extensions (or the infinite number not shown here) with information only from the original spacetime. Observe that even if one gets rid of the second asymptotically flat part and cuts out the spacetime by the blue line that represents a possible surface of the radiating body (a star say) removing all that is to the left of that blue line, the two extensions are still different beyond FEH, one keeps a radiating star and the one on the left does not.}
\label{Vaidya2}
\end{figure}

In \cite{FMS} a constructive procedure was put forward to achieve reasonable extensions of (\ref{Vaidya}), equivalent to those of Israel, if one only knows $m(u)$ for $-\infty < u <\infty$. The following properties were proven to be needed for the extensions
$$
\mu \neq 0, \hspace{1cm} \frac{d^k m}{du^k} (u\rightarrow \infty) =0 \hspace{3mm} {\rm for all} \hspace{2mm} k\in \mathbb{N}
$$
plus the {\em extended} mass function must have a critical point, a local minimum, at the null hypersurface $r=2\mu$. There exist an infinite number of ways to prolong $m(u)$ keeping these characteristics, and the only realistic way to find a proper one is to consider the physics of the underlying situation, say by knowing things about the star that emits the radiation. Imagine, for instance, that one arrives at $r=2\mu$ with the A3H becoming null there, as is actually the case represented in figure \ref{Vaidya1}. Among the infinite possible ways to extend the metric, there are two obvious choices which are totally inequivalent: to assume that the mass function remains constant in the extended part, or to add a mirror symmetrical spacetime. In both cases one obtains smooth spacetimes. These two inequivalent possibilities are represented in figure \ref{Vaidya2}. And another remark: the spacetime on the left in figure \ref{Vaidya2} is yet another extension of the Schwarzschild solution (\ref{Schw})!

To summarize, the incompleteness proven by Penrose'e theorem may just be due to extendibility of the spacetime, and the extensions may turn the original Cauchy hypersurface into just a partial one, therefore a {\em Cauchy horizon} arises, see \cite{HE,S1,Wald,P5}. The choice of spacetime beyond the Cauchy horizon suffers from all the problems just discussed with regard extensions, so that the final global spacetime may be regular or not. A possible way out to this unsatisfactory situation would be that generic Cauchy horizons turned out to be unstable, such as the one in Kerr's black hole \cite{DL,PoI}. As an aside remark, to this day there is no theorem ``predicting'' the ring singularity in the analitycally maximally extended Kerr black hole \cite{GP}, see however \cite{L,Mi,Mi0}.

\section{Closing remarks}
The singularity theorems provide supporting evidence for the need to better understand the behaviour of the gravitational field well inside black holes and probably at the initial stages of our Universe \cite{BV2,VW}. Whether or not this has to be resolved by a quantum theory of gravity is unclear \cite{Boj,Wall}, but most probably some kind of corrections to GR, or some violation of classical convergence conditions, will probably be required to restore global regularity.
The important problem of extensions of incomplete spacetimes should eventually be clarified by taking decisions on when and how extensions should be performed, and under which criteria. 

Letting that aside, I hope this expository paper will convey my belief that Penrose's ideas and, in particular his 1965 milestone paper \cite{P}, have influenced the evolution of GR and the understanding of the gravitational field in many different directions and in a revolutionary manner. In particular, Penrose deserved the Nobel prize for his gigantic contributions to GR, for the advancements that his results provoked in the understanding of black holes and, perhaps more importantly, for his visionary, unusually keen foresighted, writings. As an emblematic example, let me quote him once more \cite{P4}

\begin{quote}
Does it
follow, then, that nothing of very great astrophysical interest is likely to arise
out of collapse? Do we merely deduce the existence of a few additional dark
“objects” which do little else but contribute, slightly, to the overall mass density
of the universe? Or might it be that such “objects”, while themselves hidden from
direct observation, could play some sort of catalytic role in producing observable
effects on a much larger scale. The “seeding” of galaxies is one possibility which
springs to mind. And if “black holes” are born of violent events, might they not
occasionally be ejected with high velocities when such events occur! (The one
thing we can be sure about is that they would hold together!) I do not really want
to make any very specific suggestions here.
I only wish to make a plea for “black holes” to
be taken seriously and their consequences to
be explored in full detail. For who is to say,
without careful study, that they cannot play
some important part in the shaping of
observed phenomena?
\end{quote}

Let me finish with some personal remarks speaking as a member of a truly joyful relativity
community that, lately, is starting to `get used' to such distinctions as the Nobel prizes. The half 2020 Nobel prize has a particularly special significance to us, because it awarded mainly mathematical work. This constitutes an unequivocal recognition of {\em mathematical relativity} and of its \underline{physical implications}. The 2020 Nobel prize was a very pleasant surprise that heartened a theoretical community who, figuratively speaking, felt somehow “represented” in the award.

\bmhead{Acknowledgments}
This article is freely based on my contribution {\it The 1965 singularity theorem and its legacy} to the conference ``Singularity theorems, causality, and all that: A tribute to Roger Penrose'' (SCRI21), June 14--18, 2021. I am grateful to the organizers for giving me the opportunity to deliver the opening talk.

\bmhead{Funding}
Supported under Grant No. FIS2017-85076-P (Spanish MINECO/AEI/FEDER, EU).



\bigskip
%
%
%
%
%





\bibliography{sn-bibliography}


\end{document}